\documentclass[preprint,3p,sort&compress,twocolumn]{elsarticle}

\newcommand{\bra}[1]{\langle #1 |}
\newcommand{\ket}[1]{| #1 \rangle}
\newcommand{\ie}{\textit{i.e. }}
\newcommand{\eq}[1]{Eq. (\ref{eq:#1})}
\newcommand{\eqs}[2]{Eqs. (\ref{eq:#1}) and (\ref{eq:#2})}
\newcommand{\expect}[1]{\langle #1 \rangle}
\newcommand{\overlap}[2]{\langle #1 | #2 \rangle}

\usepackage{graphicx}
\usepackage{amsmath}
\usepackage{amssymb}
\usepackage{amsfonts}

\journal{Chemical Physics Letters}

\begin{document}

\title{Electronic correlations in organometallic complexes}

\author{A. C.~Jacko\corref{cor1}}\cortext[cor1]{Corresponding Author}
\ead{jacko@physics.uq.edu.au}
\author{B. J.~Powell\corref{cor2}}

\address{Centre for Organic Photonics and Electronics, School of Mathematics and Physics, \\ The University of Queensland, St Lucia, 4072, Australia}

\date{\today}

\begin{abstract}
We investigate an effective model for organometallic complexes (with potential uses in optoelectronic devices) via both exact diagonalisation and the configuration interaction singles (CIS) approximation.
This model captures a number of important features of organometallic complexes, notably the sensitivity of the radiative decay rate to small chemical changes.
We find that for large parameter ranges the CIS approximation accurately reproduces the low energy excitations and hence the photophysical properties of the exact solution. This suggests that electronic correlations do \emph{not} play an important role in these complexes.
This explains why time-dependent density functional theory works surprisingly well in these complexes.
\end{abstract}


\maketitle

\section{Introduction}

Organometallic complexes have significant potential for use as the optically active materials in  organic photovoltaic (OPV) devices \cite{hagfeldt00,dimitrakopoulos02,fernandez08}, organic light emitting diodes (OLEDs) \cite{friend99,forrest04,li05,lo06}, and organic light emitting field effect transistors (OLEFETs) \cite{cicoira07,namdas09}. However, progress in this field has been hampered by difficulties in identifying useful design rules for new materials. The optoelectronic functionality of such complexes depends on the properties of their lowest few excited states. Therefore, in order to aid with the design of the future generation of organometallic complexes, one might think that models are needed to accurately describe excited states on a case-by-case basis. However, an alternative approach is to introduce models that can describe whole classes of materials and thereby understand and predict trends across a range of potential complexes \cite{jacko10b}. 

The transition metal cores; typically a iridium, ruthenium, platinum or palladium atom; of the organometallic complexes discussed in this letter are associated with strong electronic correlations. The electrons in these materials are strongly correlated because of their tight confinement in $d$ and $f$ atomic orbitals, which increases the relative magnitude of electronic interactions \cite{schofield99,dagotto05} and can have significant effects on the properties of these materials \cite{hewsonbook,ghosh06,jacko09}.
A powerful approach to modelling strongly correlated electrons is to introduce effective Hamiltonians via a physically-motivated reduction in the many-body basis.  Effective Hamiltonians  have played a crucial role in explaining the strong correlation effects in both inorganic \cite{jacko09,hewsonbook,kouwenhoven01} and organic materials \cite{kouwenhoven01,powell10,jacko09} and have provided new insights in to some organometallic compounds  \cite{hewsonbook,kouwenhoven01,labute02,labute04}. Thus far such models have not been widely used to model the optical excitations of organometallic complexes with potential for OPV, OLED or OLEFET applications \cite{kober82,haneder08,jacko10a,jacko10b}. To date the modelling of these complexes has mostly been based on first principles quantum chemistry calculations.  Effective Hamiltonian and first principles approaches each have their own set of advantages and drawbacks and combining the two approaches is often extremely powerful \cite{jacko10b,kotliar06}. 

Because of the relatively large size of the complexes of interest for OPV, OLED or OLEFET applications most first principles studies are limited to density functional theory (DFT) and time-dependent DFT (TDDFT). One puzzle is that TD(DFT) approaches have been shown to give reasonably accurate predictions for a number of organometalic compounds \cite{smith11a,hay02,obara06,rusanova06,bomben09, wilson10, baccouche10, butschke10, mendes10, jacobsen10}, which one would not expect if their are strong electronic correlations \cite{fulde,powell09book}. 
The complementary approach of constructing and solving effective model Hamiltonians can address some of the shortcomings of TDDFT. One of the advantages of effective Hamiltonians is that they can be used test the validity of approximations and investigate the parameter regimes where they break down. By reducing the model to the few degrees of freedom needed to describe low energy processes, one can include a more detailed description of electronic correlations, at the cost of some chemical specificity. On reducing the description of the system to only the effective low energy interactions, one often finds that the same model is applicable to many seemingly disparate systems. For example, the Anderson single impurity model has been applied to such varied systems as magnetic impurities in metals, quantum dots in semiconductor heterostructures, carbon nanotubes, and single molecule transistors \cite{hewsonbook,kouwenhoven01}, the Heisenberg model has found applications from magnetic systems to quantum computers to valence bond theories of chemistry \cite{powell09book}   and the Schrieffer-Su-Heeger model has been used to understand the properties of a variety of polymers \cite{su80,heeger88}. 

Here we investigate an effective Hamiltonian for heteroleptic organometallic complexes via both exact diagonalisation of the Hamiltonian (full configuration interaction) and the configuration interaction singles (CIS) approximation based on in a Hartree--Fock reference ground state. We show that, for most of the parameter space of the model, CIS is an extremely accurate approximation. We give criteria for when the CIS approximation breaks down and show that CIS can accurately predict a number of trends observed in  both the exact solution of the model and in experiments on organometallic complexes, such as large changes in the radiative rate from small changes in model parameters, corresponding to subtle chemical substitutions on the organic ligand.

\section{Methods}

We study a model Hamiltonian for complexes where one metal orbital and one pair of frontier ligand orbitals dominate the low energy physics \cite{jacko10a} (this is observed in a wide variety of heteroleptic complexes for example Ru(NH$_3$)$_2$Cl$_2$(bqdi) \cite{rusanova06}, Ru(dcbpy)(bpy)$_2$ \cite{bomben09}, or Pt(cnpmic) \cite{haneder08}, further, it has been argued that solvent effects or excited state geometry relaxation can cause one ligand to dominate the physics of homoleptic complexes such as Ir(ppy)$_3$ \cite{Hofbeck}):
\begin{eqnarray} \label{eq:thehamiltonian}
\hat{H} &=& \sum_{\sigma} \bigg[ \varepsilon   \hat{M}^\dagger_{\sigma} \hat{M}_{\sigma} + \Delta  \hat{L}^\dagger_{\sigma} \hat{L}_{\sigma}  \nonumber \\
&&\hspace{0.7cm}+ t^{H} \left(\hat{H}^\dagger_{\sigma} \hat{M}_{\sigma} + \hat{M}^\dagger_{\sigma} \hat{H}_{\sigma} \right) \nonumber \\ &&
\hspace{0.7cm} + t^{L} \left(\hat{L}^\dagger_{\sigma} \hat{M}_{\sigma} + \hat{M}^\dagger_{\sigma} \hat{L}_{\sigma} \right) \nonumber \\
&& \hspace{0.7cm}+ \sum_{\sigma'} \Big( V_{HL} \hat{n}_{H\sigma} \hat{n}_{L\sigma'}  \nonumber \\ && 
\hspace{1.8cm}+ V_{HM} \hat{n}_{H\sigma} \hat{n}_{M\sigma'}  \nonumber \\ &&
\hspace{1.8cm}+ V_{LM} \hat{n}_{L\sigma} \hat{n}_{M\sigma'}  \Big) \bigg]  \nonumber \\ && 
 + U_H \hat{n}_{H\uparrow} \hat{n}_{H\downarrow} + U_L \hat{n}_{L\uparrow} \hat{n}_{L\downarrow} \nonumber \\ && + U_M \hat{n}_{M\uparrow} \hat{n}_{M\downarrow}- J \vec{S}_{H} \cdot \vec{S}_{L},
\end{eqnarray}
where 
$\hat{n}_H \equiv \sum_{\sigma} \hat{H}^\dagger_{\sigma} \hat{H}_{\sigma}$,
$\hat{n}_L \equiv \sum_{\sigma} \hat{L}^\dagger_{\sigma} \hat{L}_{\sigma}$,
$\hat{n}_M \equiv \sum_{\sigma} \hat{M}^\dagger_{\sigma} \hat{M}_{\sigma}$,
$\vec{S}_H = \sum_{\alpha, \beta} \hat{H}_\alpha^\dagger \vec{\sigma}_{\alpha \beta} \hat{H}_\beta$,
$\vec{S}_L = \sum_{\alpha, \beta} \hat{L}_\alpha^\dagger \vec{\sigma}_{\alpha \beta} \hat{L}_\beta$,
$\hat{H}^{(\dagger)}_{\sigma}$ annihilates (creates) an electron in the `renormalized HOMO level of the ligand',
$\hat{L}^{(\dagger)}_{\sigma}$ annihilates (creates) an electron in the `renormalized LUMO level of the ligand',
$\hat{M}^{(\dagger)}_{\sigma}$ annihilates (creates) an electron in the `renormalized metal orbital',
$\varepsilon$ is the difference in the renormalized energies of the metal orbital and the ligand HOMO, $\Delta$ is the renormalized HOMO-LUMO gap, $t^{H}$ ($t^{L}$) is the effective hopping amplitude between the metal orbital and the ligand HOMO (LUMO), $J$ is the effective exchange interaction between the ligand HOMO and the ligand LUMO, $U_i$ is the effective Coulomb repulsion between two electrons in orbital $i$ ($i \in \{H, L, M \}$), and $V_{ij}$ is the effective Coulomb repulsion between an electron in orbital $i$ and another in orbital $j$. The single electron orbitals  are chosen to be orthonormal, i.e., $\{\hat{H}_{\sigma},\hat{M}_{\sigma}\}=\{\hat{H}_{\sigma},\hat{L}_{\sigma}\}=\{\hat{M}_{\sigma},\hat{L}_{\sigma}\}=0$. For the discussion of the solution of this model, below, it will be helpful to define the state $|vac\rangle$, the `vacuum' state with $\langle \hat{n}_H\rangle=\langle \hat{n}_M\rangle=\langle \hat{n}_L\rangle=0$.

It is important to note that renormalized frontier (HOMO, LUMO and metal) orbitals may be quite different from the bare HOMO, LUMO and metal orbitals. 
  The state $\hat{H}^\dagger_{\sigma}|vac\rangle$ ($\hat{L}^\dagger_{\sigma}|vac\rangle$) will, in general, be significantly different from the highest occupied (lowest unoccupied) molecular orbital of the ligand found via a Hartree--Fock calculation on an isolated ligand, and similarly the `metal orbital' might have electron density on the ligands \cite{jacko10a,jacko10b,powell09book}. This means that the second quantized notation we adopt in this letter is significantly more convenient than wavefunction based notations.
  
The set of interactions retained in Hamiltonian (\ref{eq:thehamiltonian}) is equivalent to an INDO (incomplete neglect of differential overlap) Hamiltonian \cite{fulde}, in that we have included all direct Coloumb interactions, but only the local exchange interactions.
Like the one electron orbitals, the Hamiltonian parameters are renormalized.  In principle one could derive the parameters by integrating out high energy degrees of freedom from first principles calculations \cite{scriven09,gunnarssonbook,brocks04,canocortes07,scriven09B}. However, this is a highly computationally demanding task therefore, a more fruitful approach is to make use of semi-emprical parameterisations \cite{powell09book}. 

To find the Hartree--Fock ground state of this Hamiltonian one constructs a variational single Slater determinant state and minimizes the energy of that state \cite{foresman92}.
To construct the single Slater determinant, we first chose, without loss of generality, an arbitrary set of three mutually orthogonal one-hole operators, corresponding to linear superpositions of the three orbitals in this model
\begin{eqnarray}
\hat{a}_\sigma &\equiv& \sin \phi \left[ \sin \theta \hat{H}_\sigma + \cos \theta \hat{M}_\sigma \right] \notag \\&& + \cos\phi \hat{L}_\sigma,\label{eq:holea}  \\
\hat{b}_\sigma &\equiv& \sin \beta \left( \cos \phi \big[ \sin \theta \hat{H}_\sigma + \cos \theta \hat{M}_\sigma \right] \notag \\&& \hspace{1cm}
- \sin\phi \hat{L}_\sigma \Big) \nonumber \\ && 
+ \cos\beta \Big(\cos \theta \hat{H}_\sigma - \sin \theta \hat{M}_\sigma \Big),\label{eq:holeb}\\
\hat{c}_\sigma &\equiv& \cos \beta \Big( \cos \phi \big[ \sin \theta \hat{H}_\sigma + \cos \theta \hat{M}_\sigma \big] \notag \\&&  \hspace{1cm}
- \sin\phi \hat{L}_\sigma \Big)  \nonumber \\ &&  - \sin\beta \Big(\cos \theta \hat{H}_\sigma - \sin \theta \hat{M}_\sigma \Big).\label{eq:holec}
\end{eqnarray}
These states are orthonormal for any choice of the angles $\theta, \phi$ and $\beta$, e.g. $\{\hat{a}_\sigma,\hat{b}_\sigma\}=\{\hat{b}_\sigma,\hat{c}_\sigma\}=\{\hat{c}_\sigma,\hat{a}_\sigma\}=0$ (since, by definition, $\hat{H}$, $\hat{M}$ and $\hat{L}$ act on orthogonal spaces).

We construct a closed shell four electron single-determinant ground state from one of these arbitrary states, writing it as two holes in the six electron state
\begin{eqnarray}
\ket{S_{0}^{HF}} &\equiv& \hat{a}_\downarrow \hat{a}_\uparrow \left(\hat{L}^\dagger_\uparrow \hat{L}^\dagger_\downarrow \hat{M}^\dagger_\uparrow \hat{M}^\dagger_\downarrow \hat{H}^\dagger_\uparrow \hat{H}^\dagger_\downarrow \right)\ket{vac} \nonumber \\
 &\equiv& \hat{a}_\downarrow \hat{a}_\uparrow \ket{\Phi_6} \label{eq:mfgs} \end{eqnarray} \begin{eqnarray}
 &=& 
 \cos^2\phi   \ket{0}  
 +  \sin^2\theta  \sin^2\phi \hat{L}^\dagger_\uparrow \hat{L}^\dagger_\downarrow \hat{H}_\downarrow \hat{H}_\uparrow \ket{0}  \nonumber \\ &&
 + \cos^2\theta  \sin^2\phi  \hat{L}^\dagger_\uparrow \hat{L}^\dagger_\downarrow \hat{M}_\downarrow \hat{M}_\uparrow \ket{0} \nonumber \\ &&
 + \sqrt{2} \cos\theta \cos\phi  \sin\phi   \nonumber \\ &&\qquad \times\frac{1}{\sqrt{2}}(\hat{L}^\dagger_\uparrow \hat{M}_\uparrow + \hat{L}^\dagger_\downarrow \hat{M}_\downarrow) \ket{0}  \nonumber \\ &&
  + \sqrt{2}  \sin\theta  \cos\phi \sin\phi   \nonumber \\ && \qquad \times\frac{1}{\sqrt{2}}(\hat{L}^\dagger_\uparrow \hat{H}_\uparrow + \hat{L}^\dagger_\downarrow \hat{H}_\downarrow) \ket{0} \nonumber \\ &&
    + \sqrt{2} \cos\theta  \sin\theta  \sin^2\phi   \nonumber \\ && \qquad  \times \frac{1}{\sqrt{2}}\hat{L}^\dagger_\uparrow \hat{L}^\dagger_\downarrow ( \hat{H}_\uparrow \hat{M}_\downarrow - \hat{H}_\downarrow \hat{M}_\uparrow) \ket{0} \label{eq:mfgsexp} \end{eqnarray}
    \begin{eqnarray}
     &\equiv &  \cos^2\phi   \ket{0}  
     +  \sin^2\theta  \sin^2\phi  \ket{^1 LC^2}   \nonumber \\ &&
     + \cos^2\theta  \sin^2\phi  \ket{^1 MLCT^2}  \nonumber \\ &&
     + \sqrt{2} \cos\theta \cos\phi  \sin\phi  \ket{^1 MLCT^1}     \nonumber \\ &&
     + \sqrt{2}  \sin\theta  \cos\phi \sin\phi  \ket{^1 LC^1}    \nonumber \\ &&
     + \sqrt{2} \cos\theta  \sin\theta  \sin^2\phi \ket{^1 MH^2}, \label{eq:mfgsexpbasis}
\end{eqnarray}
where $\ket{0}\equiv \hat{M}^\dagger_\uparrow \hat{M}^\dagger_\downarrow \hat{H}^\dagger_\uparrow \hat{H}^\dagger_\downarrow \ket{vac} = \hat{L}_\downarrow \hat{L}_\uparrow \ket{\Phi_6}$ and $\ket{\Phi_6}$ is the state with 6 electrons, \ie every orbital doubly occupied. It is worth briefly discussing the basis in which we have expressed the variational state in \eqs{mfgsexp}{mfgsexpbasis}, as it provides a natural basis in which to study this system and we will refer to these states frequently. This basis of states comprises of the singlets defined by \eqs{mfgsexp}{mfgsexpbasis} and their triplet counterparts has $\langle \hat{n}_H\rangle, \langle \hat{n}_M\rangle$ and $\langle \hat{n}_L\rangle$ as quantum numbers. These states are the eigenstates of the Hamiltonian, \eq{thehamiltonian}, for $t^H = t^L =0$. The superscript prefix in the state label denotes spin degeneracy, and the superscript suffix denotes LUMO occupancy $\langle \hat{n}_L\rangle$. The reference state, $\ket{0}$, has no electrons in the ligand LUMO, and two each in the HOMO and metal orbitals.  $\ket{^1 LC^1}$ and $\ket{^3 LC^1}$ are, respectively, singlet and triplet ligand centred excitations, where one of the HOMO electrons of the reference state has been excited to the LUMO. In the metal-to-ligand charge transfer states, $\ket{^1MLCT^1}$ and $\ket{^3MLCT^1}$, an electron has been moved from the metal orbital to the LUMO. $\ket{^1LC^2}$ and $\ket{^1MLCT^2}$ are the states where two such ligand centred or MLCT excitations, respectively, have occurred, while $\ket{^1MH^2}$ and $\ket{^3MH^2}$ have one excitation of each kind. 

\eqs{holea}{mfgsexp} give a natural interpretation of the angles $\theta$ and $\phi$. $\phi$ determines the degree of ligand LUMO character in the ground state. For $\phi =0$ the ground state is pure reference state with $\langle \hat{n}_L\rangle = 0$, whereas for $\phi = \pi/2$ the ground state has $\langle \hat{n}_L\rangle = 2$. The angle $\theta$ determines the degree of $MLCT$ or $LC$ character in the ground state, with $\theta=0$ meaning purely $MLCT$ ($\langle \hat{n}_H\rangle = 2$) and $\theta=\pi/2$ purely $LC$ ($\langle \hat{n}_M\rangle = 2$). From \eqs{holeb}{holec} we can see that $\beta$ determines the degree of ligand LUMO versus HOMO-metal character in the $b$- and $c$-hole states.

The energy of the Hartree--Fock state is given by
\begin{eqnarray}
E^{HF}(\theta,\phi) &=& \bra{S_{0}^{HF}}\hat{H}\ket{S_{0}^{HF}} \nonumber \\
&=& \bra{\Phi_6} \hat{a}_\uparrow^\dagger \hat{a}_\downarrow^\dagger   \hat{H} \hat{a}_\downarrow \hat{a}_\uparrow \ket{\Phi_6},
\end{eqnarray}
and must be minimized with respect to $\theta$ and $\phi$. This minimum is the Hartree--Fock ground state energy $E^{HF}_0$, defined as $E^{HF}_0 \equiv E^{HF}(\theta_0,\phi_0)$.

For the CIS calculation we need to find the singly excited states. The singly excited states are constructed by replacing one of the $a$-holes with a $b$- or $c$-hole. The $S_z = \pm 1$ states are of the form
\begin{equation}
\hat{b}_\sigma \hat{a}_{\bar{\sigma}}^\dagger \ket{S_{0}^{HF}} =   \hat{b}_\sigma \hat{a}_{\bar{\sigma}}^\dagger \hat{a}_\downarrow \hat{a}_\uparrow \ket{\Phi_6}
\end{equation}
or
\begin{equation}
\hat{c}_\sigma \hat{a}_{\bar{\sigma}}^\dagger \ket{S_{0}^{HF}} =   \hat{c}_\sigma \hat{a}_{\bar{\sigma}}^\dagger \hat{a}_\downarrow \hat{a}_\uparrow \ket{\Phi_6}
\end{equation}
while the $S_z = 0$ states are of the form
\[
\frac{1}{\sqrt{2}}\left(\hat{b}_\downarrow \hat{a}_\downarrow^\dagger \pm \hat{b}_\uparrow \hat{a}_\uparrow^\dagger\right) \hat{a}_\downarrow \hat{a}_\uparrow \ket{\Phi_6} \qquad
\]
\begin{equation} \label{eq:excitedab}
\qquad = \frac{1}{\sqrt{2}}\left(\hat{b}_\downarrow \hat{a}_\uparrow \mp \hat{b}_\uparrow \hat{a}_\downarrow\right)\ket{\Phi_6},
\end{equation}
or
\[
\frac{1}{\sqrt{2}}\left(\hat{c}_\downarrow \hat{a}_\downarrow^\dagger \pm \hat{c}_\uparrow \hat{a}_\uparrow^\dagger\right) \hat{a}_\downarrow \hat{a}_\uparrow \ket{\Phi_6} \qquad
\]
\begin{equation} \label{eq:excitedac}
\qquad = \frac{1}{\sqrt{2}}\left(\hat{c}_\downarrow \hat{a}_\uparrow \mp \hat{c}_\uparrow \hat{a}_\downarrow\right)\ket{\Phi_6},
\end{equation}
which ensures that they are spin eigenstates.
The $S_z = 0$ states defined above are a combination of two Slater determinants, while the $S_z = \pm 1$ states are single Slater determinants. To perform the CIS calculation one computes the Hamiltonian matrix in this basis of singly excited states and solves it self-consistently \cite{foresman92,fetterwalecka}. 
In this case, the remaining variational angle $\beta$ allows us to directly minimize one of the singly excited states energies, equivalent to diagonalizing the CIS Hamiltonian matrix. Here we choose the singlet excited state defined in \eq{excitedab} as the lowest excited state. 

We perform the minimization of the lowest singlet and triplet states with respect to $\beta$ separately for the singlet and triplet sectors (leading to a different value of $\beta$ for each spin sector, which we denote $\beta_S$ and $\beta_T$ respectively). Thus all three free parameters in the definition of the one-hole basis states are fixed for both the singlet or triplet sectors ($[\theta_0,\phi_0,\beta_S]$ and $[\theta_0,\phi_0,\beta_T]$, respectively). Hence, the energy of every excited state can now be computed.

\section{Results}

We now explicitly compute the CIS energies and compare them to the exact eigenvalues of the model. To do this we use a standard set of parameter values around which we will vary one parameter at a time to investigate its effects. A typical set of parameters for organometallic complexes is: $J = 1$ eV, $\Delta = 3$ eV,  $\varepsilon =0.25$ eV,  $t^H = 0.1$ eV,  $t^L = 0.1$ eV,  $U_M=U_H=U_L=U= 3$ eV and  $V_{HL} = V_{HM} = V_{LM} = V= 3$ eV \cite{jacko10a}. 

\begin{figure}[f]
	\centering
		\includegraphics[width=0.45\textwidth]{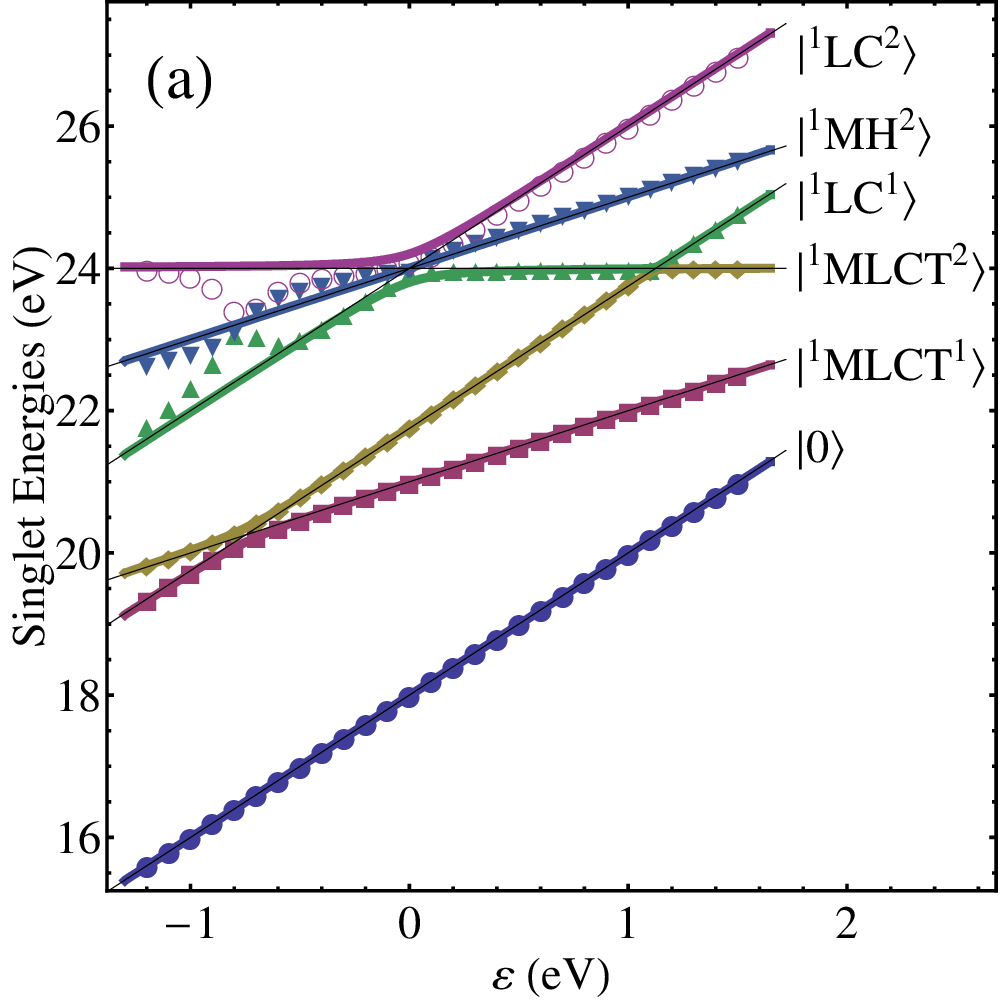} \\
		\includegraphics[width=0.45\textwidth]{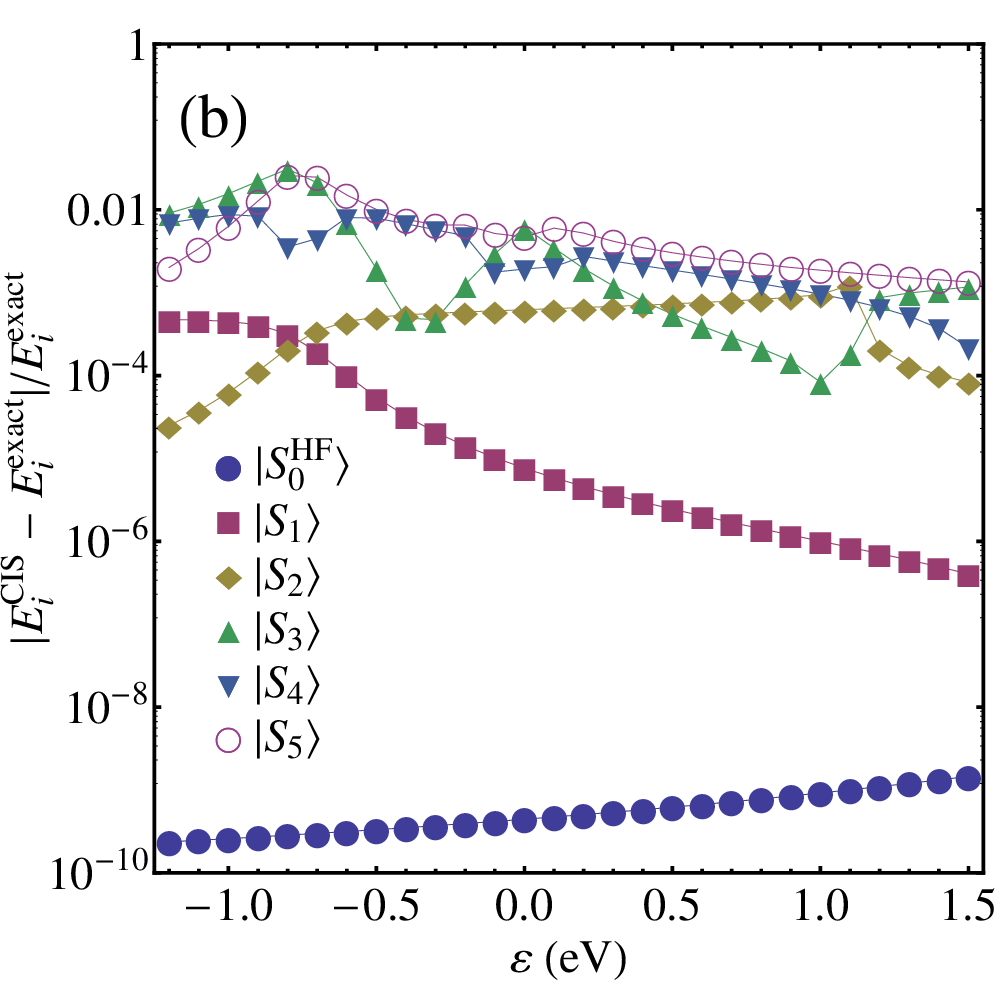}
	\caption{[Colour Online] (a) Singlet energies as a function of $\varepsilon$ and (b) the relative error in those energies on a logarithmic scale. In panel (a), the thick solid lines are the exact results, the points are the results of the CIS calculations, and the thin black lines are the $t^H=t^L=0$ energies. These figures show that the energies of the ground state and singly excited states are well approximated in this parameter regime. The doubly excited states energies are most accurate away from avoided crossings. The labels on the right hand side of (a) indicate the $t^H=t^L=0$ eigenstates. The energies of these states are all linearly dependent on $\varepsilon$, with gradient $\expect{n_M}$.}
	\label{fig:onemetaloneligand_variational_ep_sing}
\end{figure}

We find that the CIS excited states energies are typically extremely close to the exact energies. Fig. \ref{fig:onemetaloneligand_variational_ep_sing} shows that in the ground state and singly excited states the error in the energy is well below 1\%. It is also interesting to compare both the exact and CIS results to the  (exact) solutions for $t^H = t^L = 0$. In this limit the solutions are eigenstates of $\hat n_H$, $\hat n_M$ and $\hat n_L$ and thus are given by the basis defined by and below \eqs{mfgsexp}{mfgsexpbasis}.
When $t^H \neq 0 \neq t^L$ these basis states mix. Significant mixing only occurs when the energy difference between the $t^H = t^L = 0$ eigenstates is the same order as the matrix elements (proportional to $t^H$ or $t^L$) coupling them. Since $t^H$ and $t^L$ are expected to be quite small (0.1 eV in the parameterization used here), for most of parameter space the exact eigenstates are close to being pure $t^H=t^L=0$ eigenstates.

Fig. \ref{fig:onemetaloneligand_variational_ep_sing} also shows that around the avoided crossings, the error in the higher excited states energies increases (even if the crossing does not involve any of the high lying excited states, such as near $\varepsilon = -0.75$ eV). Recall that $\theta$ and $\phi$ are determined by optimizing the ground states energy, and $\beta$ is determined by optimizing the first excited states energy. Thus, once the energies of these states have been optimized, there are no more free parameters to optimize the higher excited states. Nevertheless, away from avoided crossings the CIS calculation does a remarkably good job of predicting all the the excitation energies, even those of the double excitations.

\begin{figure}
	\centering
		\includegraphics[width=0.45\textwidth]{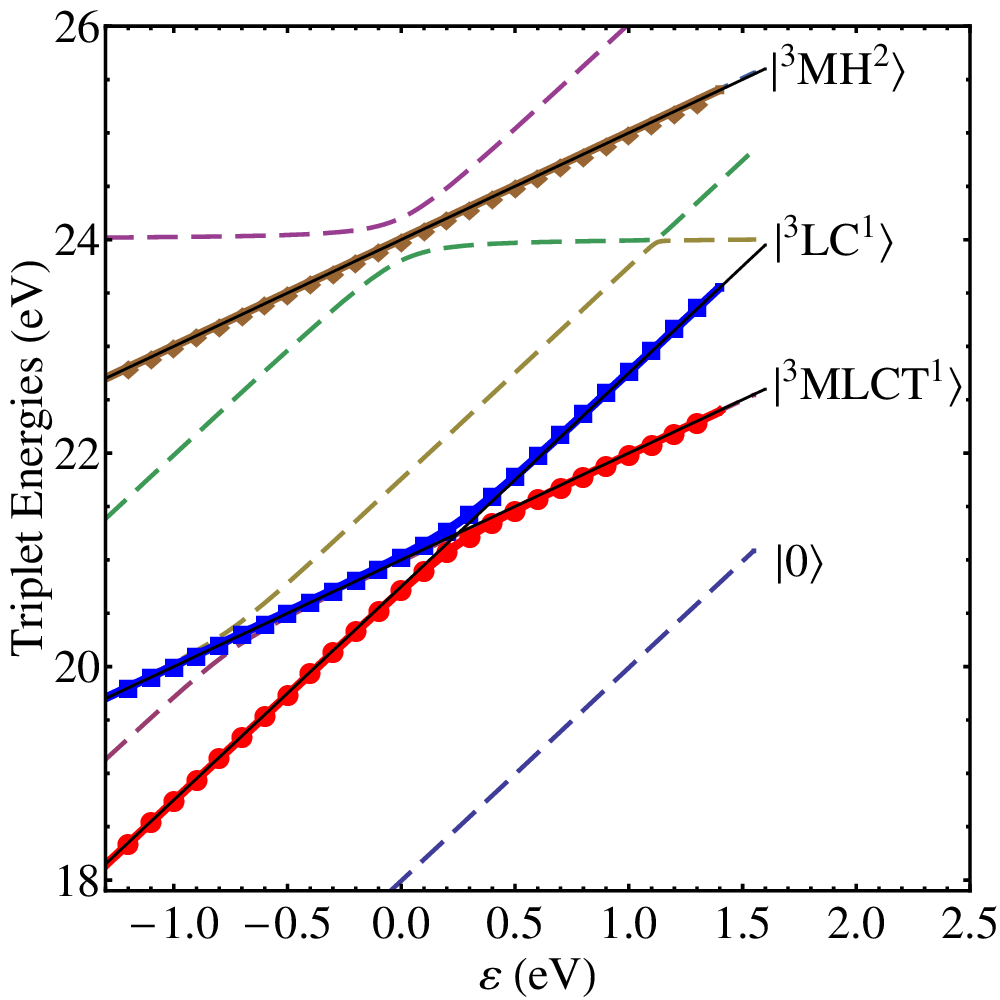}
	\caption{[Colour Online] Exact and CIS triplet energies as a function of $\varepsilon$, with the exact singlet energies show for comparison. The thick solid lines are the exact results for the triplets, the points are the results of the CIS calculations for the triplet states, the thin black lines are the $t^H=t^L=0$ triplet energies, and the dashed lines are the exact singlet results. This figure shows that for $\varepsilon > 0.25$ eV, the lowest excited triplet state is nearly degenerate with the lowest excited singlet, and both are predominantly metal-to-ligand charge transfer states; while for $-0.75 < \varepsilon < 0.25$ eV the lowest triplet is mostly $x$. The third triplet is nearly degenerate with the fourth excited singlet as both are predominantly $\ket{MH^2}$ (and the only singlet--triplet splitting in this model occurs due to mixing with the $\ket{LC^1}$ states).
	The labels on the right hand side indicate the $t^H=t^L=0$ triplet eigenstates. The energies of these states are all linearly dependent on $\varepsilon$, with gradient $\expect{n_M}$.}
	\label{fig:onemetaloneligand_variational_ep_trip}
\end{figure}

Fig. \ref{fig:onemetaloneligand_variational_ep_trip} shows that this is also true of the triplet excited states. Once the ground state energy is optimized there is just one minimization parameter left in the triplet subspace, $\beta_T$ , and yet all three triplet states are within 0.1\% of the exact energy over the range plotted here.

\begin{figure}
	\centering
		\includegraphics[width=0.45\textwidth]{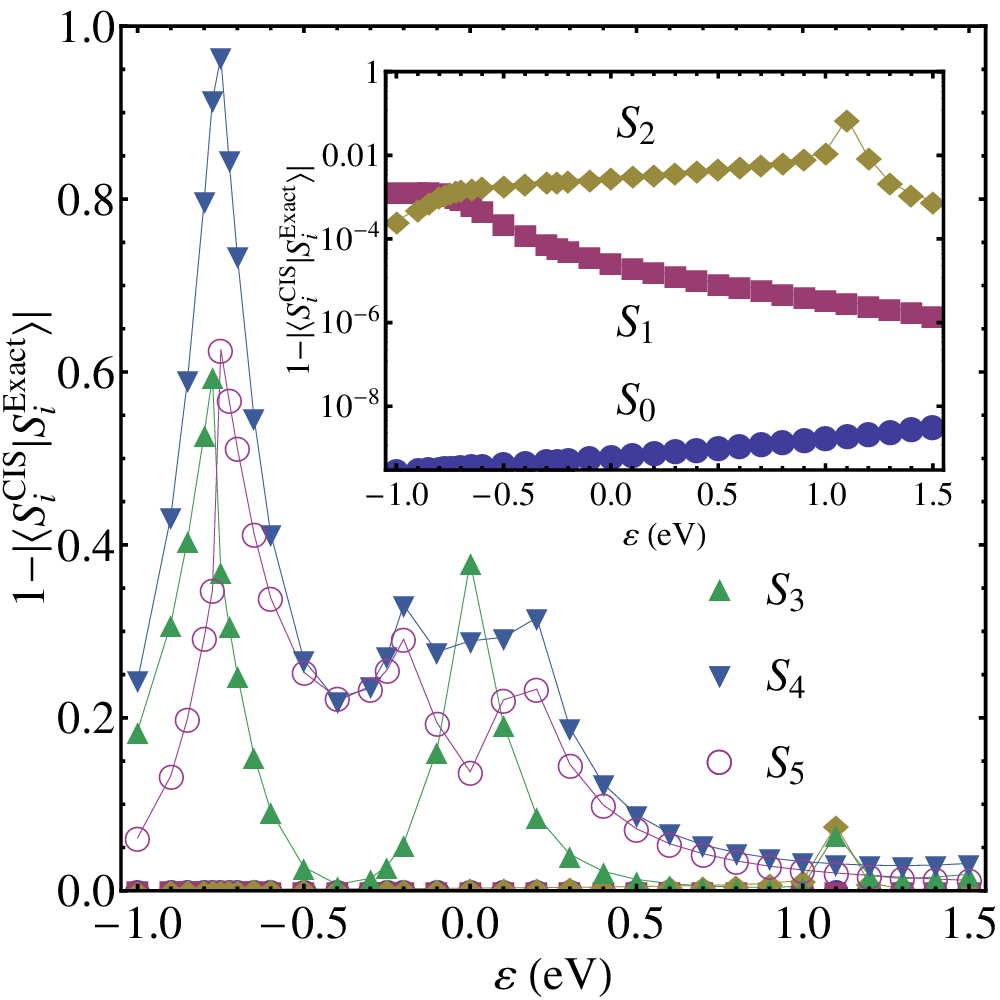}
	\caption{[Colour Online] The error in the mean field wavefunctions, defined as 1 minus the overlap between the exact and CIS states, versus $\varepsilon$. The ground state and first two excited states have errors that are typically much less than 1\% for the parameter values in this figure; the inset shows the error in these states on a logarithmic scale. The largest error is in the three doubly occupied states around $\varepsilon = -0.75$ eV, coinciding with the avoided crossing between the first two excited states (see Fig. \ref{fig:onemetaloneligand_variational_ep_sing}). The two other local maxima in the error occur around $\varepsilon =0$ eV, where the three $n_L = 2$ states have an avoided crossing, and at $\varepsilon = 1.1$ eV, where $\ket{^1 LC^1}$ and $\ket{^1 MLCT^2}$ have an avoided crossing. The errors in the wavefunctions for the ground state and first excited state are below 1\% for the parameter values in this figure. Away from $\varepsilon = 1.1$ eV, the error in the wavefunction of the second excited state is also below 1\%. From this data we conclude that the approximate ground state and singly excited states will typically do quite well, while the higher excited states only do well away from avoided crossings, that is, when they can once again be well represented by the $t^H=t^L=0$ eigenstates.}
	\label{fig:onemetaloneligand_variational_ep_sing_overlap}
\end{figure}

It is important to stress that having a good variational estimate of a states energy is a necessary but not sufficient condition for having a good variational wavefunction. Accurate wavefunctions are necessary, however, to make accurate predictions of the physical and chemical properties of the complexes under consideration. 
Fig. \ref{fig:onemetaloneligand_variational_ep_sing_overlap} shows the error in the singlet wavefunctions, defined as $1 - \left|\overlap{S_i^{CIS}}{S_i^{Exact}}\right|$. It is clear from this figure that around the avoided crossings at $\varepsilon = -0.75$ eV, $0$ eV, and $1.1$ eV the higher excited states are not correctly reproduced by the mean field CIS procedure. Fig. \ref{fig:onemetaloneligand_variational_ep_sing_overlap} also makes it clear that the singly excited states are much closer to the exact solutions than the doubly excited states.

\begin{figure}
	\centering
		\includegraphics[width=0.45\textwidth]{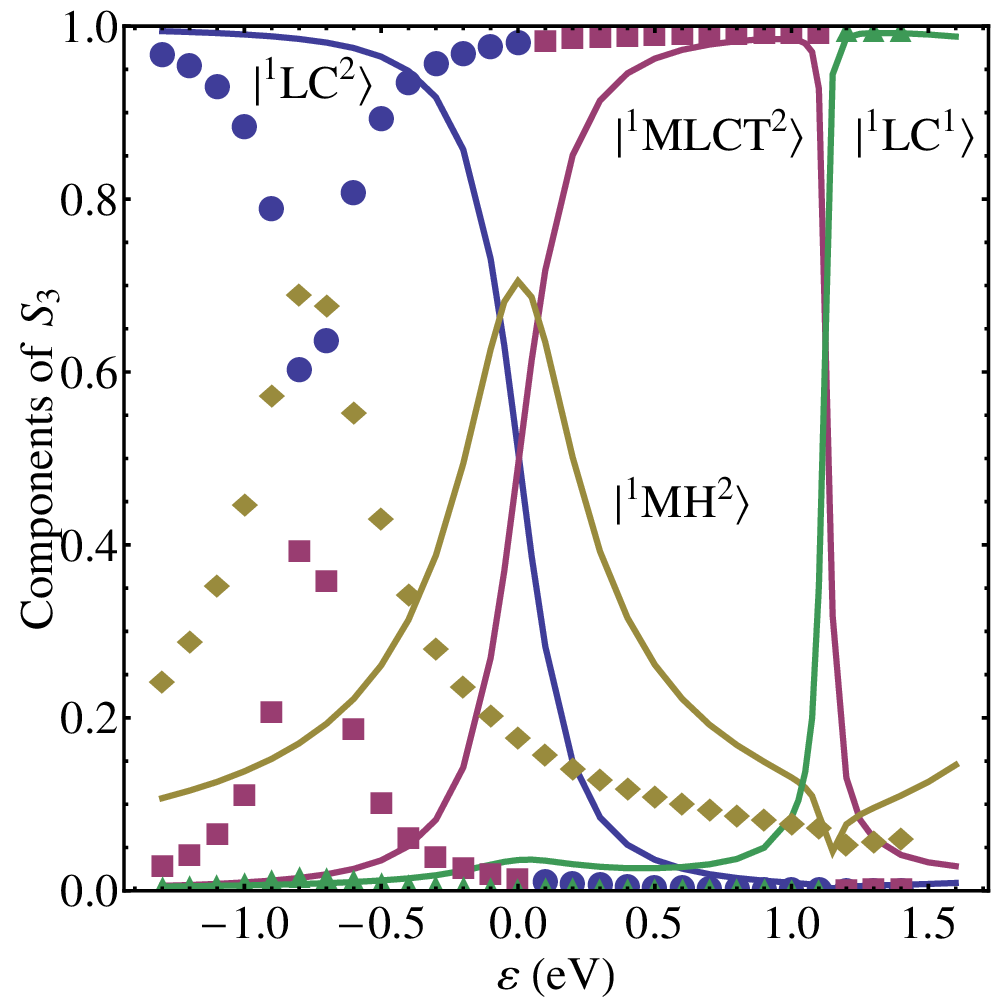}
	\caption{[Colour Online] The magnitude of the $t^H=t^L=0$ eigenstates in the exact (solid lines) and CIS (markers) third excited singlet state, $S_3$ (green triangles in Figs. \ref{fig:onemetaloneligand_variational_ep_sing} and \ref{fig:onemetaloneligand_variational_ep_sing_overlap}). The $\ket{0}$ and $\ket{^1 MLCT^1}$ components are not shown as they are not significant in the region plotted. Around $\varepsilon = 0$ there is a three-way avoided crossing, involving all three $n_L=2$ states: $\ket{^1 LC^2}$, $\ket{^1 MLCT^2}$, and  $\ket{^1 MH^2}$.  At $\varepsilon = 1.1$ eV this excited state changes character from mostly $\ket{^1 MLCT^2}$ to mostly $\ket{^1 LC^1}$. In the CIS solution these transitions are sharp crossovers rather than the gradual crossings seen in the exact solution. Around $\varepsilon = -0.75$ eV one can see the effects of avoided crossing between the $\ket{S_1}$ and $\ket{S_2}$ states (composed primarily of the $\ket{^1 LC^1}$ and $\ket{^1 MLCT^1}$ states). The CIS procedure optimizes the energy of the $\ket{S_1}$ state; at the avoided crossing where this state is not well described by a single $t^H=t^L=0$ eigenstate the CIS procedure causes the higher excited states to also be strongly mixed.}
	\label{fig:onemetaloneligand_nl2_components}
\end{figure}

Further understanding of the successes and failures of the CIS approximation can be gained by examining the wavefunctions of the excited states. For example, Fig. \ref{fig:onemetaloneligand_nl2_components} shows the dominant components of the exact and CIS solutions for the third excited singlet state, $S_3$. 
Note that, except near the avoided crossings at $\varepsilon = 0$ eV and 1.1 eV, one $t^H=t^L=0$ eigenstate dominates the exact eigenstate.
The singlet eigenstates are nearly pure $t^H=t^L=0$ eigenstates because the hopping terms $t^H$ and $t^L$ mixing these states are expected to be small compared to the other Hamiltonian parameters, which determine the energy gaps between these states. The $t^H=t^L=0$ eigenstates are spin symmetrised pairs of Slater determinants; these states have the minimum static correlation. Near avoided crossings, when these states are strongly mixed in the exact eigenstates, the exact states have strong static correlations. The CIS solution does a poor job of reproducing the exact state near the avoided crossings at $\varepsilon = -0.75 \text{ eV}, 0 \text{ eV}$ and $1.1 \text{ eV}$, as seen previously in Fig. \ref{fig:onemetaloneligand_variational_ep_sing_overlap}. In the CIS solution the avoided crossings are sharp crossovers rather than the smooth transitions seen in the exact solution.

\begin{figure}
	\centering
		\includegraphics[width=0.45\textwidth]{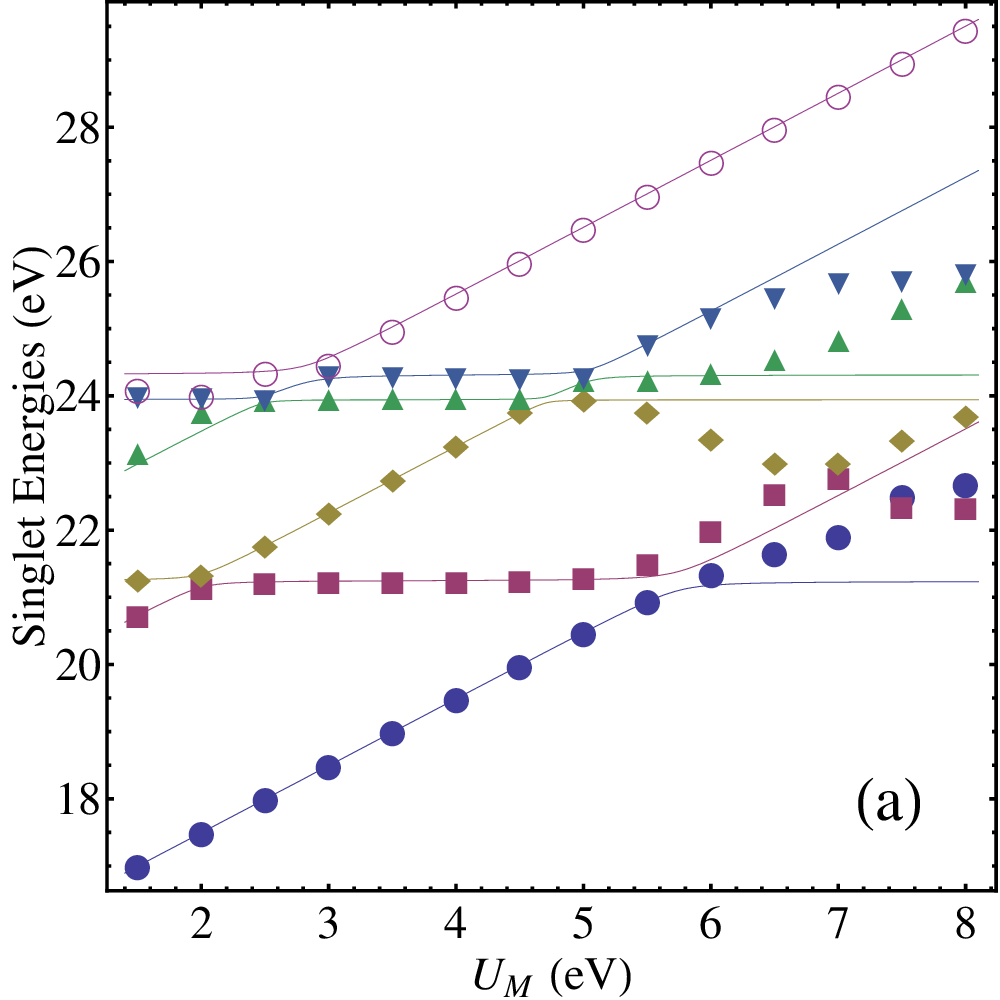} \\
		\includegraphics[width=0.45\textwidth]{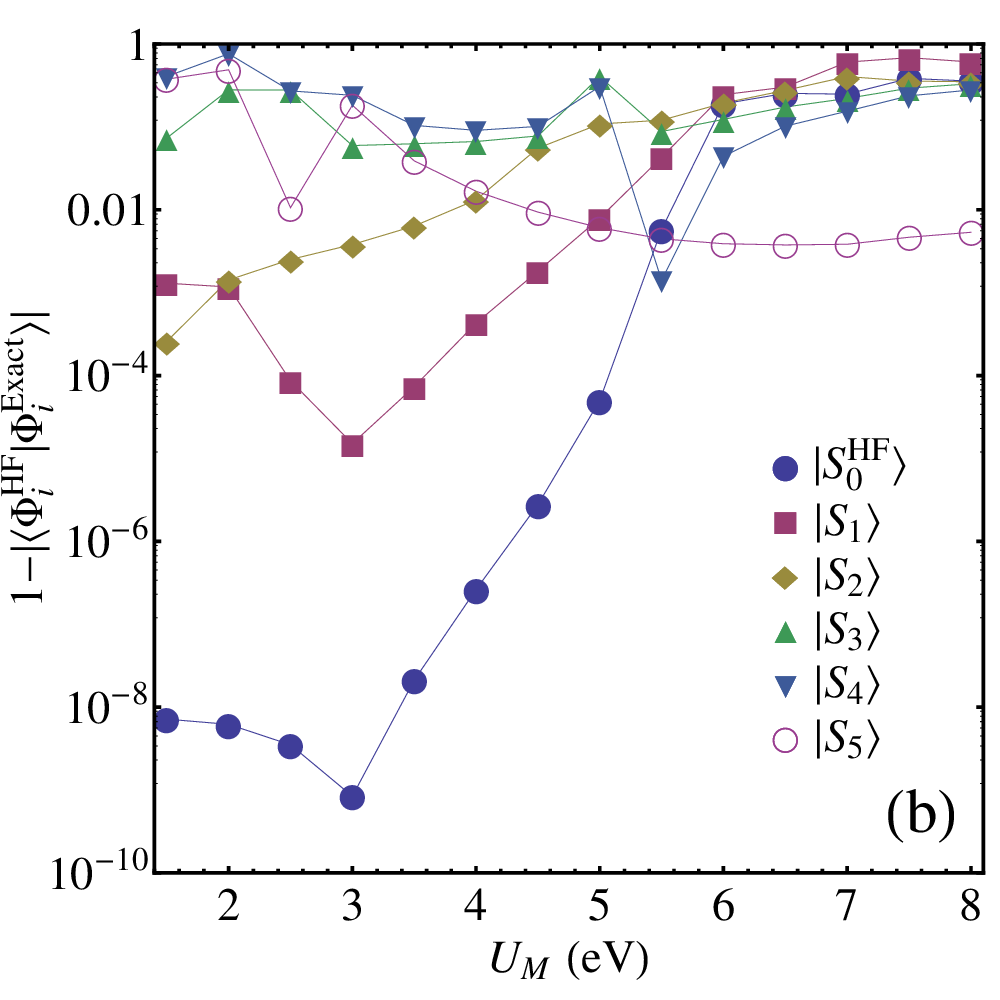}
	\caption{[Colour Online] (a) Singlet energies as a function of $U_M$, and (b) the error in the CIS wavefunctions on a logarithmic scale, over a large range of $U_M$. In (a), the coloured solid lines are the exact results, and the points are the results of the CIS calculations. It is clear that for $U_M \gtrsim 5.5$ eV, when the ground state is no longer predominantly the reference basis state, the mean field approximation breaks down. For $U_M > 5.5$ eV, the ground state contains a large $\ket{^1 MLCT^1}$ component, and therefore is no longer closed shell. In this regime electronic correlations become important and the CIS approximation fails dramatically.}
	\label{fig:onemetaloneligand_variational_um_sing_long}
\end{figure}

When constructing the variational ground state \eq{mfgs} it was assumed that the exact ground state is closed shell. We see in Fig. \ref{fig:onemetaloneligand_variational_um_sing_long} that if the Coloumb energy $U_M$ becomes large then this assumption is not valid. Above $U_M = 6$ eV the exact ground state is predominantly the open shell $\ket{^1 MLCT^1}$ state. Since there is no way to write an open shell state in terms of two $a$-holes the error in the wavefunction increases quickly, and is over 25\% for $U_M = 6.5$ eV. 
Thus, this model has a clear dividing line between a weakly correlated regime, where the ground state is closed shell, and a strongly correlated regime, where the ground state is open shell. The CIS approximation fails dramatically when in the strongly correlated regime ($U_M\gtrsim5.5$ for the parameters in Fig. \ref{fig:onemetaloneligand_variational_um_sing_long}). However, one does not expect the ground state to be open shell in organometallic complexes and thus the CIS approximation should give reliable results for these materials.

Now that we have have established the regime of validity for the CIS approximation, we can apply it to calculate other properties of the excited states with the goal of understanding the radiative properties of organometallic complexes. Of particular interest is the phosphorescent lifetime of various organometallic complexes.
Radiative emission from triplet states (phosphorescent decay) is made possible by including spin-orbit coupling. Further, it has been found that one must include spin-orbit coupling in TDDFT calculations to correctly describe the spectra of organometallic complexes \cite{smith11a}. Therefore, in order to study the radiative properties predicted by the current model we add spin-orbit coupling perturbatively to the solutions of the model Hamiltonian.
A triplet state $\ket{T_{m}}$ has a transition dipole moment, $M_{T_m}$, to the ground state, $\ket{S_0}$, that is given (to first order in the spin orbit coupling Hamiltonian, $\hat{H}_{SO}$) by
\begin{equation} \label{eq:socpert}
M_{T_m}^0 = \sum_n \frac{\bra{T_{m}} \hat{H}_{SO}\ket{S_{n}}}{E_{T_m} - E_{S_n}} M_{S_n}^0 + \frac{\bra{T_{m}} \hat{H}_{SO}\ket{S_{0}}}{E_{T_m} - E_{S_0}},
\end{equation}
where $E_{T_m}$ is the energy of the triplet state, $E_{S_n}$ is the energy of the singlet state $\ket{S_{n}}$ with transition dipole moment to the ground state $M_{S_n}^0$, the sum runs over all excited singlet states, and the final term is the direct excited state to ground state transition \cite{hochstrasser}. 

Eq. \ref{eq:socpert} is only valid if the effect of the spin-orbit coupling is perturbative, i.e., for ${\bra{T_{m}} \hat{H}_{SO}\ket{S_{n}}} \ll {E_{T_m} - E_{S_n}}$. Thus we only expect the perturbative approach to be applicable if the all singlet excited states are sufficiently different in energy from the triplet state in question.
The only singlet close in energy to the first triplet is the first excited singlet; all the other singlets are much further away in energy. In the parameter regime where the perturbative method is valid, the first excited singlet state is dominated by the $\ket{^1 MLCT^1}$ state (Fig. \ref{fig:onemetaloneligand_variational_ep_sing}) and the first triplet is dominated by $\ket{^3 LC^1}$ (Fig. \ref{fig:onemetaloneligand_variational_ep_trip}). This allows us to make the approximation that $\ket{^1 MLCT^1}$ is the only singlet state that will significantly couple to $T_1$. Further, we neglect spin-orbit coupling between states unless both have a large metal character as the magnitude of the spin-orbit coupling interaction scales as $Z^4$, where $Z$ is the atomic number. Thus, Eq. (\ref{eq:socpert}) simplifies to
\begin{eqnarray}
M_{T_1}^0 &=& M_{S_1}^0 \frac{\overlap{T_{1}}{{^3}MLCT^1}}{E_{T_1} - E_{^1 MLCT^1}} \nonumber \\
&& \times \bra{{^3}MLCT^1} \hat{H}_{SO} \ket{{^1}MLCT^1}. \label{eq:TDMT}
\end{eqnarray}

It is helpful to write the singlet eigenstates in the form
\begin{equation}
\ket{S_n} \equiv c_{S_n}^{0} \ket{0} + c_{S_n}^{^1 MLCT^1} \ket{^1 MLCT^1} + ... ,
\end{equation}
thus
\begin{equation}
M_{S_1}^0 \equiv e \bra{S_{0}}\hat{r}\ket{S_1} = e \sum_{\phi,\phi'} c_{S_1}^{\phi}c_{S_0}^{\phi'} \bra{\phi}\hat{r}\ket{\phi'},
\end{equation}
where the sums over $\phi$ and $\phi'$ run over all of the states.
In order to evaluate the transition dipole moment between $S_1$ and $S_0$ one then simply notes that  $\bra{0}\hat{r}\ket{^1 MLCT^1} \simeq 0$, due to the spatial separation between the metal and ligand orbitals, similarly other matrix elements involving metal to ligand charge transfer will be negligible.
For the typical parameters in the regime where the perturbation theory converges $\ket{S_1}\simeq\ket{^1 MLCT^1}$ and $\ket{S_0}\simeq\ket{0}$ (see Fig. \ref{fig:onemetaloneligand_variational_ep_sing}). Furthermore $c_{S_1}^{0}c_{S_0}^{0} \simeq -c_{S_1}^{^1 MLCT^1}c_{S_0}^{^1 MLCT^1}$, which can be seen numerically or derived perturbatively from the $t^L=t^H=0$ limit. Hence
\begin{equation} \label{eq:singletdipolemoment}
M_{S_1}^0 \simeq  e c_{S_1}^{0}c_{S_0}^{0} \delta \vec{r},
\end{equation}
where $\delta \vec{r} \equiv \bra{^1 MLCT^1}\hat{r}\ket{^1 MLCT^1} - \bra{0}\hat{r}\ket{0}$.
Therefore,
\begin{eqnarray}
M_{T_1}^0 &=&  e c_{S_1}^{0}c_{S_0}^{0} \frac{\overlap{T_{1}}{{^3}MLCT^1}}{E_{T_1} - E_{^1 MLCT^1}} \nonumber \\
&& \times \delta \vec{r} \bra{{^3}MLCT^1} \hat{H}_{SO} \ket{{^1}MLCT^1}.\label{eqn:tripletdipolemoment}
\end{eqnarray}

From the transition dipole moment one can find the radiative decay rate of the triplet via the Einstein $A$ coefficient,
\begin{equation} \label{eq:tripletrate}
k_R^T \equiv \frac{2 \omega^3_T}{3 \varepsilon_0 h c^3} \left|M_{T_1}^0 \right|^2 
\end{equation}
where $\varepsilon_0$ is the permittivity of free space and $\hbar \omega$ is the triplet-ground state energy gap \cite{hilborn02}. The quadratic dependence on $M_{T_1}^0$ amplifies the effects of small changes in Hamiltonian parameters on the radiative decay rate. For concreteness, we use a physically reasonable value for the strength of the spin orbit coupling, $\bra{{^3}MLCT^1} \hat{H}_{SO} \ket{{^1}MLCT^1} =$ 100 cm$^{-1}$ \cite{bersuker} and choose $\delta \vec{r}$ such that we reproduce a physically reasonable singlet radiative rate of $\sim 10^{-8}$ s$^{-1}$ (see, for example, \cite{gawelda07}).

\begin{figure}
	\centering
		\includegraphics[width=0.45\textwidth]{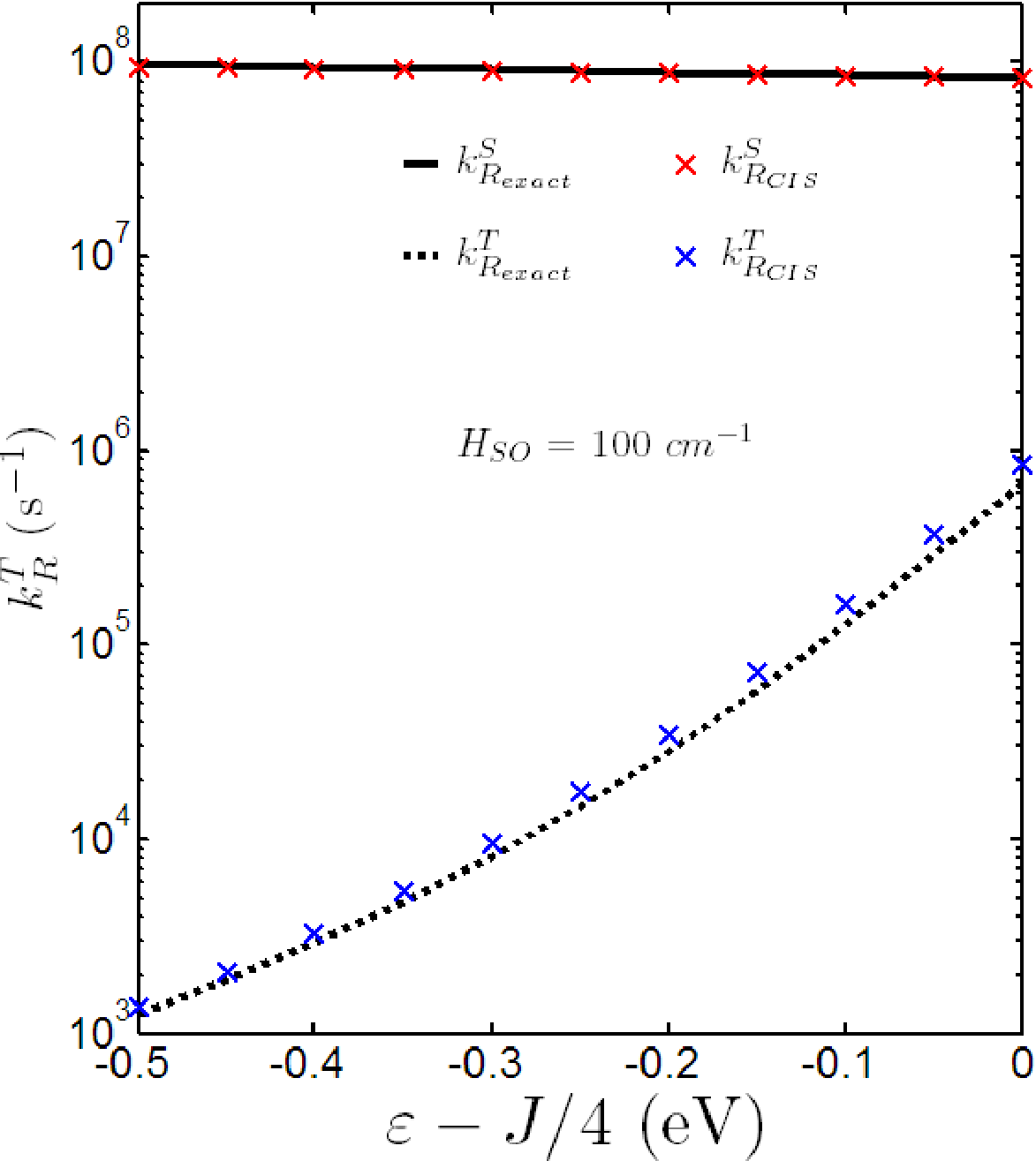}
	\caption{[Colour Online] The radiative decay rate of the lowest triplet state, as calculated by perturbation theory: in the exact solution (dotted line), and in the mean field solutions (blue crosses). The solid line is the radiative decay rate of the exact solution of the first excited singlet state, and the red crosses are the mean field solution. Note that there is no qualitative difference between the triplet radiative decay rate found from the exact and mean field solutions. The singlet radiative rate calculated from the mean field state is within 5\% of the exact rate in this figure.}
	\label{fig:onemetaloneligand_radiativerate_comparison}
\end{figure}

Fig. \ref{fig:onemetaloneligand_radiativerate_comparison} shows the CIS and exact calculations of the radiative decay rates for the lowest singlet and triplet states. The CIS approximation does a reasonable job of reproducing the exact results, although the relative errors in this experimentally observable factor are a little larger than those in the calculated energies because $k_R^T$ depends approximately on the inverse of the fourth power of energy differences.\footnote{Cf. Eqs. \ref{eqn:tripletdipolemoment} and \ref{eq:tripletrate} and note that $\ket{T_1}$ is dominated by $\ket{^3 LC^1}$ (so $E_{T_1} \simeq E_{^3 LC^1}$) with a perturbative $\ket{{^3}MLCT^1}$ component, and that $E_{^3 MLCT^1} \simeq E_{^1 MLCT^1}$, so $\overlap{{^3}MLCT^1}{T_1}\propto(E_{T_1} - E_{^1 MLCT^1})^{-1}$.} Further, for $t^H=t^L=0$  the $\ket{^3 MLCT^1}$ character of the first triplet state, and therefore its radiative rate, is exactly zero. Hence, the hybridisation of the metal and ligand orbitals is a key determining factor of the phosphorescent properties of organometallic complexes. 

There is an order of magnitude change in the triplets radiative lifetime resulting from a 0.1 eV change in $\varepsilon$, captured by both the exact and CIS solutions. This huge sensitivity of the radiative lifetime of the triplet is observed in phosphorescent complexes after small chemical modifications \cite{lo06}.   Thus, the CIS approximation captures the same physics as the exact solution; the sensitive dependence of the radiative rate on $\varepsilon$ may explain the sensitivity of organometallic complexes to small chemical changes.

\section{Discussion}

We have shown above (see particularly Fig. \ref{fig:onemetaloneligand_radiativerate_comparison}) that the model Hamiltonian investigated here predicts that small changes in the parameters lead to large changes in the  radiative lifetime. This corresponds well with what it known experimentally, where single chemical substitutions are found to lead to dramatic changes in the photo luminescent quantum yield \cite{lo06}. These changes are captured remarkably well by the CIS approximation. This suggests that the CIS approximation will be reliable and accurate as one moves to more complicated models, e.g., where more than one ligand plays an important role in the optical physics.

Given the accuracy of the CIS approximation it is interesting to briefly compare and contrast the CIS approximation with those made in TDDFT calculations. 
CIS is quite analogous to (and probably slightly worse than) the approximations typically made in TDDFT calculations.
CIS starts from a Hartree-Fock, i.e., mean field, description of the ground state. The local density approximation (LDA) is also a mean field theory, albeit a slightly different flavour. The generalized gradient approximation (GGA) and hybrid functionals include some corrections to the LDA. Therefore although DFT is in principle exact, as currently practiced, it remains close to a mean field field theory.
To find the excited states one could solve the time-dependent Kohn-Sham equations \cite{gross03}. However, it is often more convenient to make a linear response approximation, such as the random phase approximation or the Tamm-Dankoff approximation (TDA). The TDA is exactly equivalent to a configuration interaction singles (CIS) calculation \cite{foresman92}. Thus our CIS calculation using a mean-field ground state is close in both spirit and computational detail to a TDDFT calculation using the TDA. 
This means that a comparison of the CIS results from our model Hamiltonian with its exact solution provides some insight into, and perhaps and approximate lower bound on, the expected accuracy of TDDFT calculations for organometallic complexes.

\section{Conclusions}

The results presented above show that, for the lowest excited states, there are no significant differences between the exact and approximate wavefunctions and energies. This means that the key radiative properties of the complex are accurately reproduced in the CIS approximation. These results also show that there are broad ranges of parameter values for which these approximations can reproduce the energies of the higher excited states. The errors introduced by making the CIS approximation mostly effect the high energy states, which do not play a significant role in, e.g., the phosphorescence. Hence, we conclude that electronic correlations do not play an important role in determining the optoelectronic properties of these complexes.

\section*{Acknowledgments}
We are grateful to Paul Burn, Lawrence Lo, Ross McKenzie, Seth Olsen, and Arthur Smith for helpful discussions. B. J. P. was the recipient of an ARC Queen Elizabeth II Fellowship (project no. DP0878523).



\begin{thebibliography}{10}
\expandafter\ifx\csname url\endcsname\relax
  \def\url#1{\texttt{#1}}\fi
\expandafter\ifx\csname urlprefix\endcsname\relax\def\urlprefix{URL }\fi
\expandafter\ifx\csname href\endcsname\relax
  \def\href#1#2{#2} \def\path#1{#1}\fi

\bibitem{hagfeldt00}
A.~Hagfeldt, M.~Gr\"atzel, Molecular photovoltaics, Acc. Chem. Res. 33 (2000)
  269--277.

\bibitem{dimitrakopoulos02}
C.~Dimitrakopoulos, P.~Malenfant, Organic thin film transistors for large area
  electronics, Adv. Mat. 14 (2002) 99--117.

\bibitem{fernandez08}
G.~Fern{\'a}ndez, L.~S{\'a}nchez, D.~Veldman, M.~M. Wienk, C.~Atienza, D.~M.
  Guldi, R.~A.~J. Janssen, N.~Mart{\'i}n, Tetrafullerene conjugates for
  all-organic photovoltaics, J. Org. Chem. 73 (2008) 3189--3196.

\bibitem{friend99}
R.~H. Friend, R.~W. Gymer, A.~B. Holmes, J.~H. Burroughes, R.~N. Marks,
  C.~Taliani, D.~D.~C. Bradley, D.~A. {Dos Santos}, J.~L. Br\'edas,
  M.~L\"ogdlund, W.~R. Salaneck, Electroluminescence in conjugated polymers,
  Nature 397 (1999) 121--128.

\bibitem{forrest04}
S.~R. Forrest, The path to ubiquitous and low-cost organic electronic
  appliances on plastic, Nature 428 (2004) 911--918.

\bibitem{li05}
J.~Li, P.~I. Djurovich, B.~D. Alleyne, M.~Yousufuddin, N.~N. Ho, J.~C. Thomas,
  J.~C. Peters, R.~Bau, M.~E. Thompson, Synthetic control of excited-state
  properties in cyclometalated $\text{Ir(III)}$ complexes using ancillary
  ligands., Inorg. Chem. 44 (2005) 1713--1727.

\bibitem{lo06}
S.-C. Lo, C.~P. Shipley, R.~N. Bera, R.~E. Harding, A.~R. Cowley, P.~L. Burn,
  I.~D.~W. Samuel, {Blue Phosphorescence from Iridium(III) Complexes at Room
  Temperature}, Chem. Mater. 18 (2006) 5119--5129.

\bibitem{cicoira07}
F.~Cicoira, C.~Santato, Organic light emitting field effect transistors:
  Advances and perspectives, Adv. Funct. Mater. 17 (2007) 3421--3434.

\bibitem{namdas09}
E.~B. Namdas, B.~B.~Y. Hsu, Z.~Liu, S.-C. Lo, P.~L. Burn, I.~D.~W. Samuel,
  Phosphorescent light-emitting transistors: Harvesting triplet excitons, Adv.
  Mater. 21 (2009) 4957--4961.

\bibitem{jacko10b}
A.~C. Jacko, R.~H. McKenzie, B.~J. Powell, Models of organometallic complexes
  for optoelectronic applications, J. Mater. Chem. 20 (2010) 10301--10307.

\bibitem{schofield99}
A.~J. Schofield, $\text{Non-Fermi liquids}$, Contemp. Phys. 40~(2) (1999)
  95--115.

\bibitem{dagotto05}
E.~Dagotto, Complexity in strongly correlated electronic system, Science 309
  (2005) 257--262.

\bibitem{hewsonbook}
A.~C. Hewson, The Kondo Problem to Heavy Fermions, Cambridge University Press,
  1997.

\bibitem{ghosh06}
A.~Ghosh, {Transition metal spin state energetics and noninnocent systems:
  challenges for DFT in the bioinorganic arena}, J. Biol. Inorg. Chem. 11
  (2006) 712--724.

\bibitem{jacko09}
A.~C. Jacko, J.~O. Fjaerestad, B.~J. Powell, {A unified explanation of the
  Kadowaki–-Woods ratio in strongly correlated metals}, Nature Phys. 5 (2009)
  422--425.

\bibitem{kouwenhoven01}
L.~Kouwenhoven, L.~Glazman, Revival of the kondo effect, Physics World 14
  (2001) 33--38.

\bibitem{powell10}
B.~J. Powell, R.~H. McKenize, {Quantum frustration in organic Mott insulators:
  from spin liquids to unconventional superconductors}, arXiv:1007.5381 (2010).

\bibitem{labute02}
M.~X. LaBute, R.~V. Kulkarni, R.~G. Endres, D.~L. Cox, Strong electron
  correlations in cobalt valence tautomers, J. Chem. Phys. 116~(9) (2002)
  3681--3689.

\bibitem{labute04}
M.~X. LaBute, R.~G. Endres, D.~L. Cox, {An Anderson impurity model for the
  efficient sampling of adiabatic potential energy surfaces of transition metal
  complexes}, J. Chem. Phys. 121 (2004) 8221--8230.

\bibitem{kober82}
E.~M. Kober, T.~J. Meyer, {Concerning the absorption spectra of the ions
  M(bpy)$_3 ^{2+}$ (M = Fe, Ru, Os; bpy = 2,2'-bipyridine)}, Inorg. Chem. 21
  (1982) 3967--3977.

\bibitem{haneder08}
S.~Haneder, E.~D. Como, J.~Feldmann, J.~M. Lupton, C.~Lennartz, P.~Erk,
  E.~Fuchs, O.~Molt, I.~M\"unster, C.~Schildknecht, G.~Wagenblast, {Controlling
  the Radiative Rate of Deep-Blue Electrophosphorescent Organometallic
  Complexes by Singlet-Triplet Gap Engineering}, Adv. Mat. 20 (2008)
  3325--3330.

\bibitem{jacko10a}
A.~C. Jacko, B.~J. Powell, R.~H. McKenzie, Sensitivity of the photo-physical
  properties of organometallic complexes to small chemical changes, J. Chem.
  Phys. 133 (2010) 124314.

\bibitem{kotliar06}
G.~Kotliar, S.~Y. Savrasov, K.~Haule, V.~S. Oudovenko, O.~Parcollet, C.~A.
  Marianetti, Electronic structure calculations with dynamical mean-field
  theory, Rev. Mod. Phys. 78 (2006) 865.

\bibitem{smith11a}
A.~R.~G. Smith, M.~J. Riley, S.-C. Lo, P.~L. Burn, I.~R. Gentle, B.~J. Powell,
  {Relativistic effects in a phosphorescent Ir(III) complex}, Phys. Rev. B 83
  (2011) 041105(R).

\bibitem{hay02}
P.~J. Hay, Theoretical studies og the ground and excited electronic states in
  cyclometalated phenylpyridine $\text{Ir(III)}$ complexes using density
  functional theory, J. Phys. Chem. A 106 (2002) 1634--1641.

\bibitem{obara06}
S.~Obara, M.~Itabashi, F.~Okuda, S.~Tamaki, Y.~Tanabe, Y.~Ishii, K.~Nozaki,
  M.-a. Haga, {Highly Phosphorescent Iridium Complexes Containing Both
  Tridentate Bis(benzimidazolyl)-benzene or -pyridine and Bidentate
  Phenylpyridine: Synthesis, Photophysical Properties, and Theoretical Study
  of Ir-Bis(benzimidazolyl)benzene Complex}, Inorg. Chem. 45 (2006) 8907--8921.

\bibitem{rusanova06}
J.~Rusanova, E.~Rusanov, S.~I. Gorelsky, D.~Christendat, R.~Popescu, A.~A.
  Farah, R.~Beaulac, C.~Reber, A.~B.~P. Lever, The very covalent
  diammino(o-benzoquinonediimine) dichlororuthenium(ii). an example of very
  strong π-back-donation, Inorg. Chem. 45 (2006) 6246--6262.

\bibitem{bomben09}
P.~G. Bomben, K.~C.~D. Robson, P.~A. Sedach, C.~P. Berlinguette, On the
  viability of cyclometalated ru(ii) complexes for light-harvesting
  applications, Inorg. Chem. 48 (2009) 9631–9643.

\bibitem{wilson10}
G.~J. Wilson, G.~D. Will, {Density-functional analysis of the electronic
  structure of tris-bipyridyl Ru(II) sensitisers}, Inorg. Chim. Acta 363 (2010)
  1627--1638.

\bibitem{baccouche10}
A.~Baccouche, B.~Peign, F.~Ibersiene, D.~Hammoutne, A.~Boutarfaa,
  A.~Boucekkine, C.~Feuvrie, O.~Maury, I.~Ledoux, H.~L. Bozec, {Effects of the
  Metal Center and Substituting Groups on the Linear and Nonlinear Optical
  Properties of Substituted Styryl-Bipyridine Metal(II) Dichloride Complexes:
  DFT and TDDFT Computational Investigations and Harmonic Light Scattering
  Measurements}, J. Phys. Chem. A 114 (2010) 5429--5438.

\bibitem{butschke10}
B.~Butschke, S.~G. Tabrizi, H.~Schwarz, {Ion-Molecule Reactions of Rollover
  Cyclometalated [Pt(bipy-H)]$^+$ (bipy=2,2-bipyridine) with Dimethyl Ether in
  Comparison with Dimethyl Sulfide: An Experimental/Computational Study}, Chem.
  Eur. J. 16 (2010) 3962--3969.

\bibitem{mendes10}
P.~J. Mendes, T.~J. Silva, A.~P. Carvalhoa, J.~P. Ramalho, {DFT studies on
  thiophene acetylide Ru(II) complexes for nonlinear optics:
  Structure–function relationships and solvent effects}, J. Mol. Struct. 946
  (2010) 33--42.

\bibitem{jacobsen10}
H.~Jacobsen, A.~Correa, A.~Poater, C.~Costabile, L.~Cavallo, {Understanding the
  M(NHC) (NHC = N-heterocyclic carbene) bond}, Coord. Chem. Rev. 253 (2010)
  687--703.

\bibitem{fulde}
P.~Fulde, {Electron Correlations~in~Molecules~and Solids}, 3rd Edition,
  Springer, Berlin, 1995.

\bibitem{powell09book}
B.~J. Powell, {Computational Methods for Large Systems: Electronic Structure
  Approaches for Biotechnology and Nanotechnology \textit{(Ed. J. R.
  Reimers)}}, {Wiley}, 2011, Ch. {An introduction to effective low-energy
  Hamiltonians in condensed matter physics and chemistry.}

\bibitem{su80}
W.~P. Su, J.~R. Schrieffer, A.~J. Heeger, Soliton excitations in polyacetylene,
  Phys. Rev. B 22 (1980) 2099--2111.

\bibitem{heeger88}
A.~J. Heeger, S.~Kivelson, J.~R. Schrieffer, W.~P. Su, Solitons in conducting
  polymers, Rev. Mod. Phys. 60 (1988) 781--850.

\bibitem{Hofbeck}
T.~Hofbeck, H.ÊYersin, $\text{The triplet state of fac-Ir(ppy)}_3$,
  InorganicÊChemistry 49 (2010) 9290.

\bibitem{scriven09}
E.~Scriven, B.~J. Powell, {Towards the parameterisation of the Hubbard model
  for salts of BEDT-TTF: A density functional study of isolated molecules}, J.
  Chem. Phys. 130 (2009) 104508.

\bibitem{gunnarssonbook}
O.~Gunnarsson, Alkali-Doped Fullerides: Narrow-Band Solids with Unusual
  Properties, World Scientific, 2004.

\bibitem{brocks04}
G.~Brocks, J.~van~den Brink, A.~F. Morpurgo, Electronic correlations in
  oligo-acene and -thiopene organic molecular crystals, Phys. Rev. Lett. 93
  (2004) 146405.

\bibitem{canocortes07}
L.~Cano-Cort\'es, A.~Dolfen, J.~Merino, J.~Behler, B.~Delley, K.~Reuter,
  E.~Koch, {Spectral broadening due to long-range Coulomb interactions in the
  molecular metal TTF-TCNQ}, Eur. Phys. J. B 56 (2007) 173--176.

\bibitem{scriven09B}
E.~Scriven, B.~J. Powell, {Effective Coulomb interactions within BEDT-TTF
  dimers}, Phys. Rev. B 80 (2009) 205107.

\bibitem{foresman92}
J.~B. Foresman, M.~Head-Gordon, J.~A. Pople, M.~J. Frisch, Toward a systematic
  molecular orbital theory for excited states, J. Phys. Chem. 96 (1992)
  135--149.

\bibitem{fetterwalecka}
A.~L. Fetter, J.~D. Walecka, Quantum Theory of Many Particle Physics,
  McGraw-Hill, New York, 1971.

\bibitem{hochstrasser}
R.~M. Hochstrasser, Molecular Aspects of Symmetry, W. A. Benjamin, 1966.

\bibitem{hilborn02}
R.~C. Hilborn, {Einstein coefficients, cross sections, $f$ values, dipole
  moments, and all that}, arXiv:physics/0202029v1 (2002).

\bibitem{bersuker}
I.~B. Bersuker, Electronic structure and properties of transition metal
  compounds: introduction to the theory, John Wiley \& Sons, 1996.

\bibitem{gawelda07}
W.~Gawelda, A.~Cannizzo, V.-T. Pham, F.~van Mourik, C.~Bressler, M.~Chergui,
  {Ultrafast Nonadiabatic Dynamics of [FeII(bpy)$_3$]$^{2+}$ in Solution}, J.
  Am. Chem. Soc. 129 (2007) 8199--8206.

\bibitem{gross03}
M.~A.~L. Marques, E.~K.~U. Gross, {A Primer in Density Functional Theory
  \textit{Eds. C. Fiolhais, F. Nogueira, M.A.L. Marques}}, Springer Lecture
  Notes in Physics, 2003, Ch. {Time-dependent density functional theory}, pp.
  144--184.

\end{thebibliography}
\end{document}